\documentclass[epj]{webofc}
\usepackage[varg]{txfonts}   
\wocname{EPJ Web of Conferences}
\woctitle{CONF12}
\begin{document}
\selectlanguage{english}
\title{
Recent progress on QCD inputs for axion phenomenology}
%
%

\author{Claudio Bonati\inst{1,2}\fnsep\thanks{\email{claudio.bonati@df.unipi.it}
present address: Dipartimento di Fisica e Astronomia \& INFN-Sezione di Firenze, Via Sansone 1,  50019, Sesto Fiorentino (FI), Italy} \and
Massimo D'Elia\inst{1,2}\fnsep\thanks{Speaker; \email{massimo.delia@unipi.it}} \and
Marco Mariti\inst{1,2}\fnsep\thanks{\email{mariti@df.unipi.it}} \and
Guido Martinelli\inst{3,4}\fnsep\thanks{\email{guido.martinelli@roma1.infn.it}} 
\and
Michele Mesiti  \inst{1,2}\fnsep\thanks{\email{mesiti@pi.infn.it}}  
\and
Francesco Negro\inst{2}\fnsep\thanks{\email{fnegro@pi.infn.it}}  
\and
Francesco Sanfilippo\inst{5}\fnsep\thanks{\email{f.sanfilippo@soton.ac.uk}
present address: INFN, Sez. di Roma Tre, Via della Vasca Navale 84, I-00146 Rome, Italy}  
\and
Giovanni Villadoro 
\inst{6}\fnsep\thanks{\email{giovanni.villadoro@ictp.it}} 
}

\institute{Dipartimento di Fisica dell'Universit\`a di Pisa, Largo Pontecorvo 3, I-56127 Pisa, Italy 
\and
  INFN sezione di Pisa, Largo Pontecorvo 3, I-56127 Pisa, Italy       
\and
Dipartimento di Fisica dell'Universit\`a di Roma ``La Sapienza'', 
Piazzale Aldo Moro 5, I-00185 Roma, Italy \and
INFN Sezione di Roma La Sapienza, Piazzale Aldo Moro 5, I-00185 Roma, Italy \and
School of Physics and Astronomy, University of Southampton, SO17 1BJ Southampton, United Kingdom \and
Abdus Salam International Centre for Theoretical Physics, Strada Costiera 11, 34151, Trieste, Italy
}

\abstract{%
The properties of the QCD axion are strictly related to the
dependence of strong interactions on the topological parameter theta.
We present a determination of the topological properties of QCD for
temperatures up to around 600 MeV, obtained by lattice QCD simulations
with 2+1 flavors and physical quark masses. Numerical results for
the topological susceptibility, when compared to instanton gas
computations, differ both in size and in the temperature dependence.
We discuss the implications of such findings for axion
phenomenology, also in comparison to similar studies
in the literature, and the prospects for future investigations.
}
\maketitle
\section{Introduction}
\label{intro}

Axions have been advocated long 
ago~\cite{Peccei:1977hh, Peccei:1977ur, Wilczek:1977pj, Weinberg:1977ma} as a
possible solution to the so-called strong-CP problem through the Peccei-Quinn (PQ)
mechanism.  It  was soon realized that the axion could also be a possible source of dark matter~\cite{Preskill:1982cy,
Abbott:1982af, Dine:1982ah}. That puts axions among the best candidates for physics beyond the
Standard Model. 
A
reliable computation of the axion relic density and of the present axion mass
requires a quantitative
estimate of the parameters entering the axion effective potential, i.e. its 
mass and self-coupling, as a function of the temperature. 

Large part of the 
information is contained in the dependence of strong interactions
on the topological parameter $\theta$, 
which enters the pure gauge part of the QCD
Euclidean Lagrangian as
\begin{equation}\label{lagrangian}
\mathcal{L}_\theta = \frac{1}{4} F_{\mu\nu}^a(x)F_{\mu\nu}^a(x) - i
\theta q(x) \ \ ; \ \ \ 
q(x)=\frac{g^2}{64\pi^2} 
\epsilon_{\mu\nu\rho\sigma} F_{\mu\nu}^a(x) F_{\rho\sigma}^a(x)
\end{equation}
where $q(x)$ 
is the topological charge density. The
free energy density can be parametrized as follows
\begin{equation}
\label{freeene}
F(\theta,T) = F(0,T) + {1\over 2} \chi(T)
\theta^2 (1 + b_2(T) \theta^2 + b_4(T) \theta^4 + \cdots) \, ,
\end{equation}
where $\chi(T)$ is the topological susceptibility at $\theta=0$,
which is proportional to the axion mass:
\begin{equation}
\chi = \int d^4 x \langle q(x)q(0) \rangle_{\theta=0} 
= {\langle Q^2 \rangle_{\theta=0} \over {\cal V}} \label{chidef} \ ,
\end{equation}
$Q = \int d^4 x\, q(x)$ is the global topological charge 
and $\mathcal{V}=V/T$ is the 4-dimensional volume. 
The coefficients $b_n$ are proportional to the higher cumulants of the
topological charge distribution, for instance
\begin{equation}\label{eq:b2}
b_2=-\frac{\langle Q^4\rangle_{\theta=0}-
         3\langle Q^2\rangle^2_{\theta=0}}{12\langle Q^2\rangle_{\theta=0}} \, ;
\end{equation}
they provide information about axion interactions. 

Reliable analytic predictions about $\chi$ and the $b_n$ coefficients are
available only in certain regimes. For instance,
axial transformations can rotate $\theta$ to the light quark sector,
allowing to exploit chiral perturbation theory
(ChPT) in the low temperature 
regime~\cite{Weinberg:1977ma,DiVecchia:1980ve,Leutwyler:1992yt,Mao:2009sy, Guo:2015oxa,diCortona:2015ldu}, with the prediction
$\chi^{1/4}=77.8(4)$~MeV and $b_2=-0.022(1)$ in the case
of two degenerate flavors.

At high-$T$, instead, a
possible approach is based on the Dilute Instanton Gas Approximation
(DIGA). Indeed, since instantons of size
$\rho \gg 1/T$ are suppressed by thermal fluctuations, 
at asymptotically high-$T$ one can use the perturbative 1-loop estimate
for the instanton action, 
$S_{eff} \simeq 8 \pi^2 /g^2(T)$, which leads to the prediction of 
a path integral dominated by a dilute gas 
of weakly interacting objects of topological
charge one.
The $\theta$-dependence of the free energy is of the 
form~\cite{Gross:1980br,Schafer:1996wv})
\begin{equation}\label{eq:inst_gas}
F(\theta,T)-F(0,T)\simeq \chi(T)(1-\cos\theta)\ ,
\end{equation} 
and, using the 1-loop effective action, 
one obtains~\cite{Gross:1980br,Schafer:1996wv})
\begin{equation}\label{eq:chi_inst_pert}
 \chi(T) \sim T^4\left(\frac{m_l}{T}\right)^{N_f} e^{-8 \pi^2/g^2(T)}
 \sim m_l^{N_f}T^{4-\frac{11}{3}N_c-\frac{1}{3}N_f} 
\end{equation}
for the theory with $N_f$ light flavors of mass $m_l$, i.e. 
$\chi \propto T^{-7.66}$ for $N_f = 2$. Of course, the range of reliability
of the perturbative prediction is not known a priori,
and in principle 
one would trust it only for temperatures much larger than $T_c$.

A fully non-perturbative determination
of $\theta$-dependence can be obtained 
by lattice QCD simulations. 
The task is not easy: one needs to implement
a lattice discretization of the theory which is close 
enough to the continuum limit to reproduce the correct chiral
properties of fermion fields, however the correct sampling 
of topological modes is affected by critical autocorrelation
times as the continuum limit is 
approached~\cite{Alles:1996vn, DelDebbio:2002xa, DelDebbio:2004xh,
Schaefer:2010hu, Kitano:2015fla}, and is even more difficult
in the high $T$ region, where topological fluctuations become 
very rare.

In the following we present 
a lattice study of
the topological properties of $N_f = 2+1$ QCD with physical
quark masses. We adopt stout improved~\cite{Morningstar:2003gk}
staggered fermions and the tree level improved 
Symanzik action~\cite{weisz, curci}
for the pure gauge sector, exploring 
several lattice spacings,
in a range $\sim 0.05
- 0.12$~fm, and staying on a line of constant physics, which is fixed
following the determinations reported in 
Ref.~\cite{physline1, physline2}; 
the topological content of gauge configurations is extracted by 
a gauge operator after proper smoothing of gauge 
fields~\cite{cooling, Luscher:2009eq, Luscher:2010iy, Bonati:2014tqa,
Cichy:2014qta, Namekawa:2015wua, Alexandrou:2015yba}.
More technical 
details can be found in Ref.~\cite{ourpaper}.

First we consider simulations at $T = 0$, in order
to identify a proper scaling window and compare
the continuum extrapolated results for the topological 
susceptibility to the ChPT prediction. 
At finite $T$, we explore a region going up 
to around $4\, T_c$: we compare our results 
with DIGA predictions and with other recent results,
discussing implications for axion phenomenology and prospects
for future investigations.

\section{Numerical Results at Zero and Finite Temperature}
\label{sec-1}

\begin{figure}[t!]
\centering
\includegraphics[width=8cm,clip]{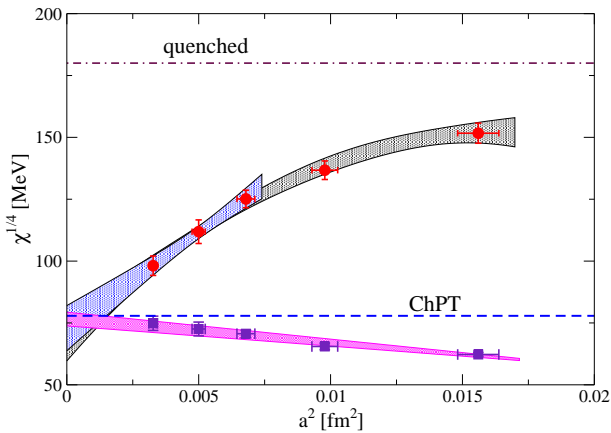}
\caption{Continuum limit of $\chi$ at $T = 0$, compared with 
the ChPT prediction and the
quenched value. Square points correspond to the quantity 
$\chi^{1/4}_{tc}(a)$  (see Eq.~\eqref{eq:tastecomb}).
}
\label{fig-1}       
\end{figure}

Results obtained for $\chi^{1/4}$ at $T = 0$, after extrapolation to 
the infinite volume limit, are reported in 
Fig.~\ref{fig-1} as a function of $a^2$, together
with the expectation from ChPT. The range of explored lattice spacings
is limited on the left by freezing of topological 
modes that we have experienced for $a~<~0.05$ fm, see Ref.~\cite{ourpaper}.

Finite cut-off effects are significant and for $a \geq 0.1$ fm
the reported values are closer to the quenched case rather than
to the ChPT prediction. Nevertheless, we can perform
a consistent continuum extrapolation of our data:
just $O(a^2)$ corrections are needed to describe results
at the three finest
lattice spacings, while $O(a^4)$ corrections must
be taken into account to describe the whole range. 
We obtain consistent continuum
extrapolated results ($\chi^{1/4} = 73(9)$ MeV in the 
former case and $\chi^{1/4} = 69(9)$ MeV in the latter),
which are also in agreement with ChPT.

The large cutoff effects are likely due to the
fact that, for those lattice spacings, our discretization fails to reproduce
the correct chiral properties of light quarks.
A typical manifestation of that, in the case of staggered quarks, is the 
fact that the full degenerate multiplet of 
(pseudo)Goldstone bosons is reproduced
only in the continuum limit, while at finite lattice spacings
one has one light pion plus other
massive states which become degenerate only as $a \to 0$.
A confirmation of that comes as one studies
the combination
\begin{equation}
\label{eq:tastecomb}
\chi^{1/4}_{tc}(a)=a\chi^{1/4}(a)\frac{m_{\pi}^{\mathrm{phys}}}{a m_{ngb}(a)}\ 
\end{equation}
where $m_{ngb}(a)$ is one of the massive pions (in particular, we adopted
the one with taste structure $\gamma_i\gamma_{\mu}$).
The values
of $\chi^{1/4}_{tc}(a)$ are reported in Fig.~\ref{fig-1} as square points:
cut-off effects are strongly reduced in this way and can be parametrized
as $O(a^2)$ corrections over the whole range.

\begin{figure}[t!]
\centering
\includegraphics[width=8cm,clip]{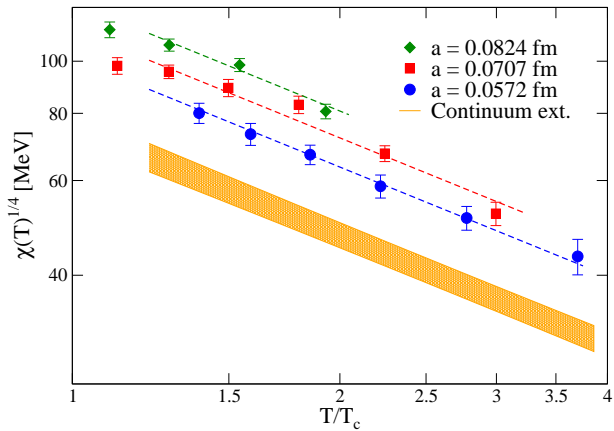}
\caption{Fourth root of the topological susceptibility as a function of 
$T$, together with the continuum extrapolation
(see Eq.~(\ref{eq:fit_susc})).
}
\label{fig-2}       
\end{figure}

Finite temperature simulations have been performed at 
the three smallest lattice
spacings, i.e.~those for which $O(a^2)$ corrections 
suffice to describe the $T = 0$ data,
and at various values of the temporal extent $N_t$, with a fixed
spatial extent $N_s = 48$, after checking the absence of appreciable
finite size effects.
Results for $\chi^{1/4}$ are 
reported in Fig.~\ref{fig-2} 
as a function of $T/T_c$ ($T_c = 155$ MeV) and have been fitted 
according to the following ansatz
\begin{equation}\label{eq:fit_susc}
\chi^{1/4}(a,T)=A_0(1+A_1 a^2)\left(\frac{T}{T_c}\right)^{A_2}\ ,
\end{equation}
which is inspired to the DIGA prediction and takes into account
also finite $a$ corrections. If one takes into account 
the range $T > 1.2\, T_c$ the ansatz works quite well 
($\chi^2/\mathrm{dof} \simeq 0.7$), however we obtain
values for the exponent $A_2$ which are significantly lower
(by roughly a factor 2) than the instanton gas prediction.

\begin{figure}[t!]
\centering
\includegraphics[width=8cm,clip]{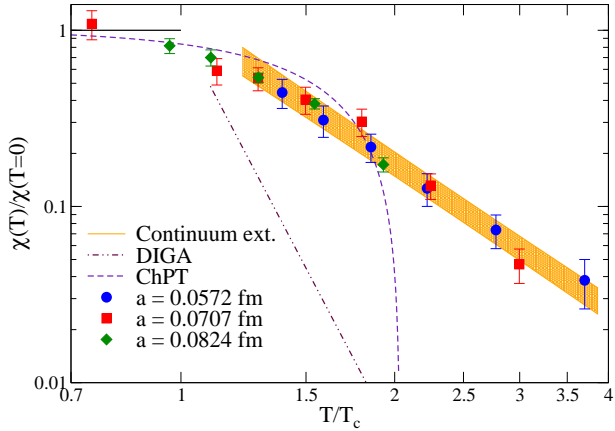}
\caption{Topological susceptibility, divided by $\chi(T=0)$
at the same lattice spacing,
as a function of $T$.
The solid line fixes the zero temperature value,
the dashed line is the finite $T$ prediction from ChPT, while
the dashed-dotted line corresponds to the slope predicted by
the dilute instanton gas prediction. 
The continuum extrapolation has been performed according to 
Eq.~(\ref{eq:fit_suscratio}).
}
\label{fig-3}       
\end{figure}

\begin{figure}[t!]
\centering
\includegraphics[width=8cm,clip]{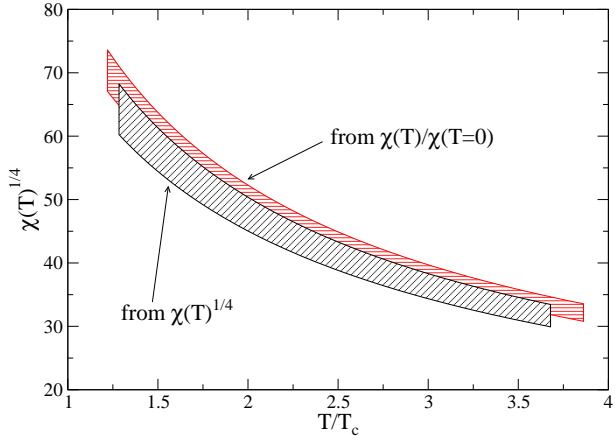}
\caption{Comparison of the two different 
continuum extrapolations for $\chi(T)^{1/4}$
($\chi(0)$ has been fixed to the ChPT prediction).
}
\label{fig-4}       
\end{figure}

Results emerge more clearly as one considers the ratio
$\chi(T,a)/\chi(T=0,a)$, which is reported in Fig.~\ref{fig-3},
together with the DIGA and ChPT predictions.
Lattice artifacts seem to be strongly suppressed in this case,
indeed the dependence on the lattice spacing is
quite mild. The difference with the DIGA prediction
is evident, indeed a best fit of our data to the function 
\begin{equation}\label{eq:fit_suscratio}
\frac{\chi(a,T)}{\chi(a,T=0)}=D_0(1+D_1 a^2) \left(\frac{T}{T_c}\right)^{D_2}\ 
\end{equation}
yields a slope coefficient $D_2 \sim - 3$, against a DIGA 
prediction $D_2 \sim -8$. On the other hand, the two different
continuum extrapolations, Fig.~\ref{fig-2} and Fig.~\ref{fig-3}, lead
to perfectly consistent results, which are compared in Fig.~\ref{fig-4}.

\subsection{Implications for Axion Cosmology}
\label{sec-2}

Whereas many high energy models exist which introduce the 
PQ symmetry and its breaking, 
the common simplest form of the low energy effective lagrangian involving
the axion field can be written as follows:
    \begin{equation*}
    \mathcal{L}_{eff}
=\mathcal{L}_{QCD}+\frac{1}{2}\partial_{\mu} a\partial^{\mu}a
    +\left(\theta+\frac{a(x)}{f_a}\right) \frac{g^2}{32 \pi^2} G \tilde G + \dots
     \end{equation*}
where $a$ is the axion field acquiring a non-zero vacuum expectation
value $\langle a \rangle$ 
after breaking of the PQ symmetry. It has only derivative 
terms, apart from a coupling to the QCD topological charge density,
which is regulated by the constant $f_a$. 
This coupling links $\theta$ dependence and 
axion phenomenology: indeed one has 
a minimum for the axion effective potential
corresponding to a zero effective $\theta$; moreover, 
assuming that $f_a$ is very large, $a$ can be considered
as quasi-static and the mass and interactions of the 
axion can be derived from the $\theta$-dependence of QCD.
For instance, one has the relation
\begin{equation}
m_a^2 (T) = \frac{\chi (T)}{f_a^2} 
\end{equation}
which would give the value of the axion mass today (from the value
of $\chi$ at  $T = 0$)
once $f_a$ is fixed.

In order to fix $f_a$, one needs further inputs. A possibility is to relate
the present abundance of axions to the amount of observed dark matter.
One of the best candidates for cosmological axion production
is misalignement: at the time the PQ symmetry
breaks, the axion field is not aligned along the minimum, 
$\theta_{eff} = 0$; the resulting excitation is
converted into axion production. 

A precise description needs the solution of the axion field evolution
which, in the static limit and taking 
into account only terms quadratic in $\theta$
in the axion potential, reads as follows
\begin{equation} \label{eq:evo}
\ddot{a}(t)+3H(t)\dot{a}(t)+m_a^2(t)a(t)=0 
\end{equation}
where $H$ is the Hubble constant. Since $H$ is a decreasing function
of time, while $m_a$ is increasing (assuming that the Universe temperature
$T$ decreases with $t$), Eq.~(\ref{eq:evo}) predicts a damped
solution as long as the friction
term wins over the pull from the potential, and becomes
oscillatory after the Universe cools below a 
temperature $T_{osc}$ for which 
$m_a (T_{osc}) \approx 3 H (T_{osc}) $. 
Shortly after that time, 
the solution develops an adiabatic invariant corresponding
to the total number of axions.

\begin{figure}[t!]
\centering
\includegraphics[width=8cm,clip]{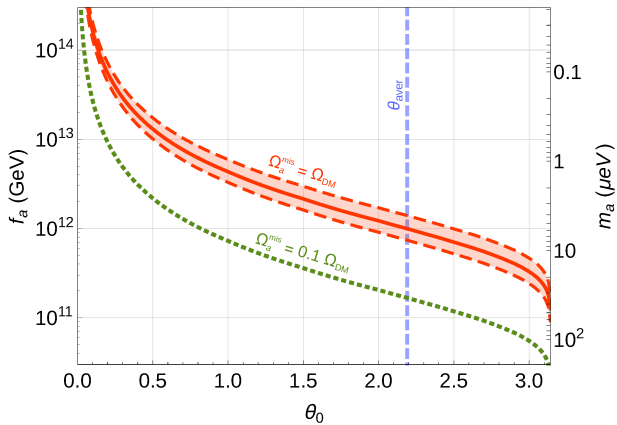}
\caption{Axion decay constant $f_a$ as a
function of the angle $\theta_0=a_0/f_a$, such that the 
misalignment contribution provides one tenth or the whole observed dark
matter abundance. The vertical dashed line fixes the average
angle $\theta_{aver}$ for the case in which the PQ
symmetry is broken after inflation.
}
\label{fig-6}       
\end{figure}

The value
of $T_{osc}$ depends both on $\chi(T)$ and on $f_a$, so that,
by requiring a given amount of axion relics, one can fix the value 
of $f_a$. Actually, an unknown constant is the initial misalignement
angle, $\theta_0$: that takes a fixed value throughout
the Universe if the PQ symmetry breaking takes place before inflation,
otherwise it can acquire
all possible values within the visible
horizon and one needs to integrate over them.
In Fig.~\ref{fig-6}, based on the results obtained
for $\chi(T)$ from our numerical simulations, we show our prediction
(see Ref.~\cite{ourpaper} for more details)
for $f_a$ and for the present value of the axion mass, depending 
on the different possible assumptions, i.e. pre- or post-inflationary
scenario and relative amount of dark matter density $\Omega_{DM}$
due to axions from the misalignment mechanism.

Our results lead to values of $f_a$ which are tipically 
one order of magnitude larger than the ones obtained from the 
dilute instanton gas model, hence to smaller values of 
the present axion mass.  This is mostly due to the different
behavior of $\chi(T)$, which shows, at least in the explored
range, a much milder dependence
on $T$ with respect to DIGA predictions. 
This result of course might be affected by 
residual systematic effects, in particular we have 
results at three different lattice spacings only in a region
which goes up to $T \sim 300$ MeV, while the oscillation 
temperature deriving from our data is of the order of few GeVs,
so that we are strongly relying on an extrapolation.
The main difficulty in approaching higher temperatures
is given by the critical slowing down observed for 
small lattice spacings and by the fact that topological
fluctuations become very rare at high $T$.

Recent results from different lattice investigations
have shown contrasting results. While the results of 
Ref.~\cite{Trunin:2015yda} are in qualitative agreement 
with ours, other studies have reported a much better agreement
with the instanton gas prediction even in the 
region right 
above $T_c$~\cite{Petreczky:2016vrs, Borsanyi:2016ksw, Taniguchi:2016tjc}.

In particular, the authors of Ref.~\cite{Borsanyi:2016ksw} manage
to determine $\chi(T)$ for $T$ up to a few GeVs and observe a slope
consistent with DIGA shortly after $T_c$. The main differences 
consist in a reweighting procedure adopted in Ref.~\cite{Borsanyi:2016ksw},
based on the lowest lying Dirac operator eigenvalues, and in a new strategy
to approach the high $T$ regime (see also Ref.~\cite{Frison:2016vuc}). 
This approach avoids direct simulations at high $T$, 
and is based instead on simulations at fixed topology, which are 
used to determine the relative weight, in the path integral, of the 
topological sectors $Q = \pm 1$ with respect to the topological sector 0,
since those are believed to be the only ones relevant 
at high $T$. Of course such a  strategy assumes right from the beginning
that the instanton gas is dilute enough and non-interacting.

\begin{figure}[t!]
\centering
\includegraphics[width=8cm,clip]{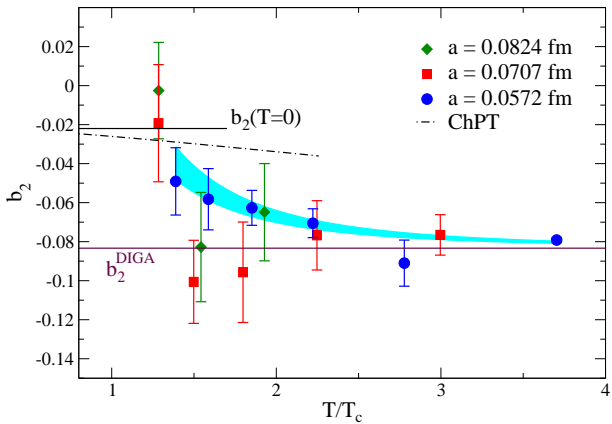}
\caption{$b_2$ as a function of $T$, together with the 
ChPT prediction at $T = 0$ ($-0.022$) and the DIGA
prediction ($-1/12$).
The light blue band is the result of a fit to the smallest lattice spacing data
using a virial expansion (see Ref.~\cite{ourpaper} for more details).}
\label{fig-5}       
\end{figure}

Good quantities to check for the diluteness hypothesis are the 
$b_n$ coefficients appearing in Eq.~(\ref{freeene}),
which are fixed by the fact that, for a non-interacting instanton gas,
one has $F(\theta) \propto (1 - \cos \theta)$.
Our results for $b_2$ are reported in Fig.~\ref{fig-5}:
even if the statistical uncertainties are still large, one sees
that deviations from the dilute gas hypothesis could be still
appreciable for $T$ up to $2-3\ T_c$.
This is in contrast with the pure gauge 
case~\cite{Bonati:2013tt, Bonati:2015uga,Bonati:2015sqt,Bonati:2016tvi},
where corrections become negligible shortly after $T_c$.
That confirms the non-triviality of fermionic contributions
and claims for future studies, which should further check
existing results. Various improvements are possible, especially
to deal with the critical slowing down of topological
modes by means of improved algorithms~\cite{deForcrand:1997fm,slab,openbc,surfing, endres}.

\vspace{1cm}
\noindent
{\large \bf Acknowledgments}
\\

Numerical simulations have been performed on the Blue Gene/Q
machine Fermi at CINECA, based on an agreement between INFN and CINECA
(under project INFN-NPQCD). FN acknowledges financial support from 
the INFN SUMA project.
Work partially supported 
by the ERC-2011 NEWPHYSICSHPC Grant Agreement Number: 27975,
by the ERC-2010 DaMESyFla Grant Agreement Number: 267985, 
 and by the
MIUR (Italy) under a contract PRIN10-11 protocollo 2010YJ2NYW.

\end{document}